# Weak Coupling of Diffusional and Phonon-like Modes in Liquids Revealed by Dynamic Kapitza Length

Tao Chen,[1] and Puqing Jiang [1, *]

[1]*School of Energy and Power Engineering, Huazhong University of Science and Technology, Wuhan, Hubei 430074, China*

Understanding heat transfer across solid-liquid interfaces is central to thermal management and energy technologies, yet whether the interfacial thermal conductance (ITC) depends on the timescale of heating remains unclear. Here we use square-pulsed source thermoreflectance, which combines time-resolved detection with broadband modulation, to probe Al-water and Al-octane interfaces. We observe a reproducible increase of the apparent ITC with modulation frequency. A control Al-silica interface shows no measurable frequency dependence, indicating that the effect is specific to liquids rather than a generic feature shared by all amorphous materials. We explain the data with a two-channel liquid picture in which diffusional and phonon-like modes exchange energy weakly over a finite nonequilibrium length. From the relative magnitude of the thermal penetration depth and the nonequilibrium length, we identify three transport regimes. These findings challenge the common assumption of fully equilibrated liquid modes and provide experimental constraints for modeling dynamic energy exchange at liquid interfaces.

Heat must cross boundaries between different phases whenever we cool electronics [1,2], store energy [3,4], or control temperature in biological and chemical systems [5,6]. At a solid-liquid boundary, that exchange of energy is governed by the interfacial thermal conductance (ITC), the proportionality between heat flux and the small temperature drop across the interface. A convenient way to express the same resistance is the Kapitza length [7], the thickness of liquid that would produce an equivalent thermal drop. Although these concepts are standard in heat transport, the microscopic pathways by which energy leaves a solid and enters a liquid remain actively debated because liquids host both collective vibrations and stochastic molecular motions [8,9].

Pump-probe thermoreflectance techniques, such as time-domain thermoreflectance (TDTR) and frequency-domain thermoreflectance (FDTR), have enabled quantitative measurements of ITC for many metal-liquid pairs and have established typical values under transient conditions [10,11]. Yet an important question has remained open: does the ITC depend on the timescale of the thermal drive? In solids, frequency-dependent responses often reveal hidden degrees of freedom and couplings between them. In liquids, where structural disorder, molecular rearrangements, and short-lived collective vibrations coexist, one might expect an even richer nonequilibrium behavior. However, the modulation frequency of the pump beam in TDTR is typically limited to above ~0.1 MHz, with most studies employing frequencies around 10 MHz. In contrast, FDTR allows for broad frequency tuning but lacks time-resolved signals [12]; these limitations obscure possible dynamical effects at solid-liquid boundaries.

Here we address this gap with a square-pulsed source (SPS) thermoreflectance approach that combines broadband frequency control with time-resolved detection [13-15] (see Supplemental Material Sec. 1.1 for details of the experimental setup and heat-transfer model). In SPS, the pump heating is square-wave modulated and the full periodic waveform is extracted synchronously using periodic waveform analysis (PWA), enabling narrow-band detection and high dynamic reserve. This broadband capability is essential for accessing the kilohertz regime while maintaining high SNR, where conventional TDTR implementations are often limited by pulse-accumulation effects and low-frequency technical noise. A detailed comparison between SPS and TDTR is provided in Supplemental Material Sec. 1.2. Using simple glass/Al/liquid stacks to minimize materials uncertainty, we examine two canonical liquids, water with strong intermolecular interactions, and octane, a non-hydrogen-bonded hydrocarbon, to test whether any dynamical behavior is specific to hydrogen bonding or instead reflects more general liquid physics.

In this Letter, we show that the ITC at metal-liquid boundaries is frequency dependent. The effect is reproducible across measurement modalities and liquids, and it can be rationalized by a minimal two-channel picture for the liquid: a diffusional channel representing slow, rearrangement-dominated motion and a phonon-like channel representing faster, vibration-dominated motion. The two channels exchange energy only weakly, so a small temperature difference can persist between them over a characteristic nonequilibrium length in the liquid. As the thermal penetration depth of the drive sweeps through this length scale, the apparent ITC evolves in a sigmoidal manner that collapses onto a single scaling curve for both water and octane, revealing three transport regimes from near-equilibrium to strongly decoupled behavior.

*Contact author: jpq2021@hust.edu.cn

These results revise a common assumption, namely that interfacial contributions from different liquid modes simply add as if perfectly equilibrated [16,17]. By establishing that the coupling between liquid modes is finite and small, our measurements provide direct experimental constraints for multi-channel descriptions of cross-phase energy transfer. Beyond clarifying a long-standing simplification, the work links interfacial heat flow to intrinsic liquid dynamics and offers a framework for predicting how interfaces will respond across timescales relevant to microelectronics cooling, energy technologies, and soft-matter systems.

The test structure consisted of water/50 nm Al/glass, where both the water and glass were treated as semi-infinite. This simple structure minimizes the propagation of uncertainties from known parameters to the ITC to be determined. Glass was selected as the substrate due to its low thermal conductivity, which promotes heat flow toward the liquid side and reduces measurement errors associated with the thermal properties of water. The Al film was deposited on the glass by electron beam evaporation, providing precise control over film thickness, which was further verified using a stylus profilometer. The thermal conductivity of the Al film was determined from its electrical resistivity measured by the four-point probe method [18], combined with the Wiedemann-Franz law.

Measurements were first performed without water, using 10 modulation frequencies spanning 1.5 kHz to 10 MHz (see Supplemental Material, Fig. S3, for experimental signals and fitting). From fitting the data, we extracted the thermal conductivity of Al and both the thermal conductivity and volumetric heat capacity of the glass substrate. The thermal conductivity of Al was found to be 72.5 W m$^{-1}$ K$^{-1}$, in good agreement with the four-point probe measurement, with the slight enhancement attributable to phonon contributions. For the glass substrate, the thermal conductivity was determined as $0.85 \pm 0.04$ W m$^{-1}$ K$^{-1}$, consistent with the manufacturer's specification (0.81-0.93 W m$^{-1}$ K$^{-1}$), while the volumetric heat capacity was $1.813 \pm 0.09$ MJ m$^{-3}$ K$^{-1}$, also in close agreement with the specified value of 1.883 MJ m$^{-3}$ K$^{-1}$.

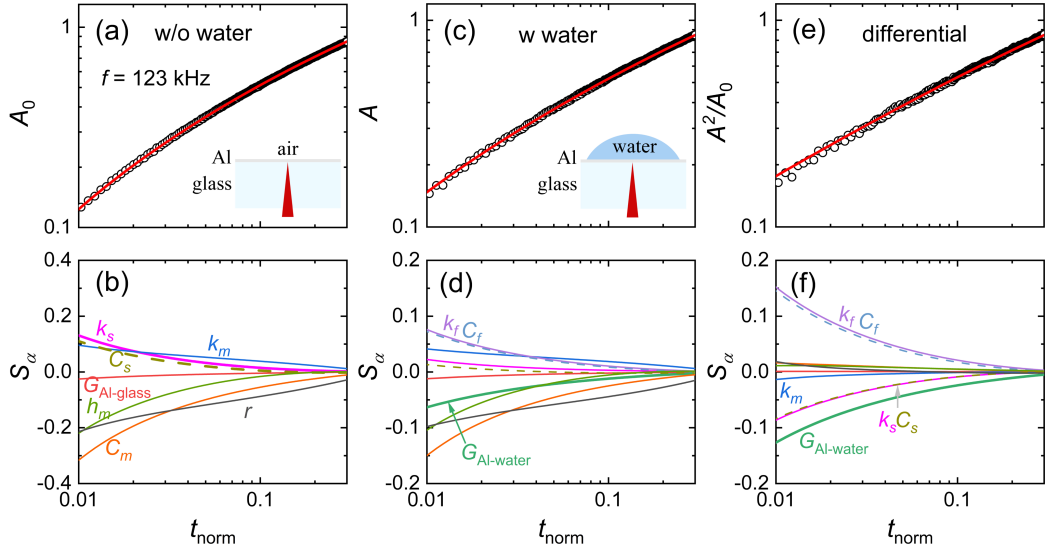

FIG. 1. Measured signals and sensitivity analysis for the differential measurement. (a) Measured signal $A_0$ without water at a modulation frequency of 123 kHz and spot radius of 13.5 μm. (b) Corresponding sensitivity coefficients for key parameters, including Al thermal conductivity ($k_m$), Al thickness ($h_m$), Al volumetric heat capacity ($C_m$), glass thermal conductivity ($k_s$), glass volumetric heat capacity ($C_s$), spot radius ($r$), and Al-glass ITC ($G_{\text{Al-glass}}$). (c) Measured signal $A$ after adding water under the same conditions. (d) Sensitivity coefficients showing additional sensitivity to water thermal properties ($k_f C_f$) and Al-water ITC ($G_{\text{Al-water}}$). (e) Squared signal ratio $A^2/A_0$. (f) Corresponding sensitivity coefficients for $A^2/A_0$, highlighting the simplified dependence on $k_f C_f$, $k_s C_s$, and $G_{\text{Al-water}}$, with reduced influence of other parameters.

Subsequently, water was added to the sample for measurement. This approach effectively performs a differential measurement, thereby suppressing the influence of known parameter uncertainties on the extracted ITC [19]. As an illustration, Figs. 1(a) and 1(b) present the measured $A_0$ signal and its sensitivity analysis for the case without water, obtained with a spot radius $r = 13.5\,\mu m$ and modulation frequency $f = 123$ kHz. The sensitivity coefficients $S_\alpha$ are defined as $S_\alpha = \partial(\ln R)/\partial(\ln \alpha)$, where $R$ is the measured signal and $\alpha$ is the parameter under analysis. In this configuration, the signal is primarily sensitive to $k_m/(C_m r^2)$, $\sqrt{k_s C_s}/(h_m C_m)$, and $G_{\text{Al-glass}}/(h_m C_m)$, where $k$ denotes thermal conductivity, $C$ the volumetric heat capacity, $h$ the film thickness, $r$ the laser spot radius, and subscripts $m$ and $s$ refer to the metal (Al) film and glass substrate, respectively. $G_{\text{Al-glass}}$ represents the Al-glass ITC.

Figures 1(c) and 1(d) show the measured signal $A$ and corresponding sensitivity analysis after adding water. Compared to Fig. 1(b), additional sensitivities to $\sqrt{k_f C_f}/(h_m C_m)$ and $G_{\text{Al-water}}/(h_m C_m)$ appear, where the subscript $f$ represents the liquid and $G_{\text{Al-water}}$ is the Al-water ITC. The overall sensitivity profiles in Figs. 1(b) and 1(d) remain similar. For example, the sensitivities to $k_m$, $h_m$, $C_m$, $r$, and $G_{\text{Al-glass}}$ in Fig. 1(b) are roughly twice those in Fig. 1(d). Figures 1(e) and 1(f) present the squared ratio $A^2/A_0$, which greatly simplifies the sensitivity, leaving dependence only on $k_f C_f$, $k_s C_s$, and $G_{\text{Al-water}}$. The uncertainty in $k_s C_s$ is estimated at ~10%, while $k_f C_f$ is taken from literature with a 5% assumed uncertainty. Propagating these uncertainties yields an overall uncertainty of 8% for $G_{\text{Al-water}}$. A detailed uncertainty analysis is provided in Sec. 2 of the Supplemental Material.

We next measured the Al-water ITC as a function of modulation frequency from 1.5 kHz to 10 MHz. The thermal conductivity and volumetric heat capacity of water were fixed at $k_f = 0.6 \pm 0.006$ W m$^{-1}$ K$^{-1}$ and $C_f = 4.18 \pm 0.08$ MJ m$^{-3}$ K$^{-1}$, respectively, while all other parameters were known except $G_{\text{Al-water}}$ (see Supplemental Material, Fig. S4, for signals and fits). At the lowest frequency (1.5 kHz), the thermal penetration depth in water, $d_p = \sqrt{k_f/(\pi f C_f)}$, is approximately 5.5 μm. To avoid full thermal penetration, the liquid layer must be thicker than $3 d_p$ [20], corresponding to ~16.5 μm. Our millimeter-thick water film easily satisfies this condition, ensuring that the extracted $G_{\text{Al-water}}$ is independent of liquid film thickness.

The measured Al-water ITCs are shown in Fig. 2(a). With increasing modulation frequency, the ITC gradually rises from 4 MW m$^{-2}$ K$^{-1}$ to 55 MW m$^{-2}$ K$^{-1}$ and saturates beyond 500 kHz. To test the universality of this dynamic behavior, we selected octane as a representative non-hydrogen-bonded liquid and extended our measurements to the Al-octane interface (see Supplemental Material, Fig. S6, for signals and fits). The Al-octane ITCs, also shown in Fig. 2(a), are markedly lower in magnitude than those of Al-water but exhibit a similar trend.

We note that liquids and amorphous solids share a unified vibration-mediated framework for thermal conductivity in broad regimes [21]. To address whether the observed frequency dependence is merely a generic feature of disordered vibrational transport, we performed additional SPS measurements on an Al-silica interface over 2.5 kHz-10 MHz (see Supplemental Material, Fig. S7, for signals and fits). For silica, the thermal properties are well established, enabling reliable extraction of the ITC alone. The result is included in Fig. 2(a). In contrast to the liquid interfaces, the Al-silica ITC shows no discernible frequency dependence within experimental uncertainty, consistent with the expectation that vibrational degrees of freedom in amorphous solids equilibrate rapidly and can be treated effectively as a single channel.

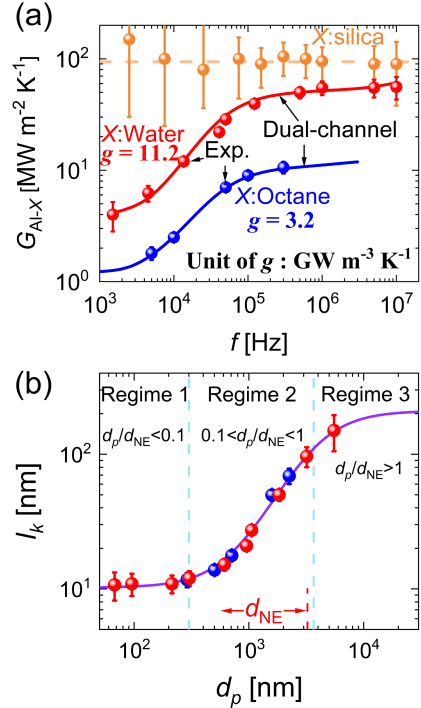

FIG. 2. (a) Measured frequency-dependent ITC of the Al-water and Al-octane interfaces. The dual-channel model reproduces the frequency dependence. Additional Al-silica control data show no measurable frequency dependence. (b) Measured Kapitza length $l_k$ ($l_k = k_f/G_{\text{Al-liquid}}$) versus thermal penetration depth $d_p$ ($d_p = \sqrt{k_f/(\pi f C_f)}$) for the Al-water and Al-octane interfaces. The sigmoidal dependence can be divided into three regimes: (i) a constant plateau at $d_p \ll d_{\text{NE}}$, (ii) a rapid increase as $d_p$ approaches $d_{\text{NE}}$, and (iii) a slower growth beyond $d_{\text{NE}}$.

To interpret the frequency dependence of the solid-liquid ITC, we adopt a dual-channel model previously applied to other frequency-dependent ITC phenomena [12,22]. A schematic of the channel division is presented in Fig. 3. On the Al side, a single channel is sufficient because electron-phonon coupling in Al is strong enough that, within our frequency range (up to 10 MHz), no measurable temperature difference develops between hypothetical channels. Indeed, single- and dual-channel representations of the glass-Al structure yield identical responses (see Supplemental Material, Sec. 3.1, including Ref. [23]).

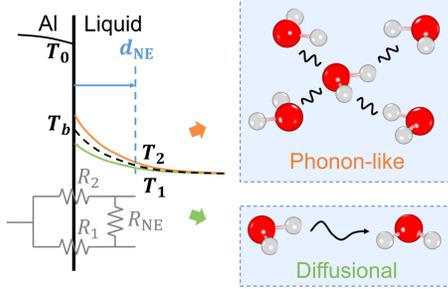

FIG. 3. Schematic illustration of the dual channel model for the Al-liquid system. The Al side is modeled as a single channel due to strong electron-phonon coupling, while the liquid side is divided into a phonon-like channel (channel 2) and a diffusional channel (channel 1), corresponding to motions above and below the Frenkel frequency ($\omega_F$), respectively. The finite coupling coefficient $g$ between the two channels leads to a temperature difference that decays exponentially with a characteristic length scale $d_{\text{NE}}$. Insets show molecular representations of the phonon-like and diffusional motions in water.

In contrast, on the liquid side we adopt a dual-channel description within the Maxwell-Frenkel viscoelastic picture [24]. Heat is carried by two channels: a diffusional channel associated with relaxational, rearrangement-mediated motion, and a phonon-like channel associated with short-lived propagating collective vibrations, which exist when the observation time is shorter than the structural relaxation time $\tau$. Here, "phonon-like" does not imply long-lived crystal phonons; rather it refers to transient collective vibrations supported by the liquid's short-range order. For water, this short-range order is evidenced by the O-O radial distribution function [25] and by orientational metrics such as the tetrahedral order parameter ($q \approx 0.6$ at ambient conditions) in experiment-based analyses [26]. We use the Frenkel frequency as a practical boundary to distinguish the two pathways, $\omega_F \sim 1/\tau$ [11].

It is important to emphasize that our analysis does not assume the experimental modulation frequency $f$ must be comparable to $\omega_F$, nor that the thermal drive "resonantly probes THz dynamics." Instead, $\omega_F$ enters only as a microscopic boundary that partitions the liquid degrees of freedom and therefore the channel-resolved thermophysical properties ($k_1, k_2, C_1, C_2$). Together with the interchannel coupling coefficient $g$, these properties set an emergent nonequilibrium length $d_{\text{NE}} = (g/k_1 + g/k_2)^{-1/2}$ [22]. In contrast, the experimentally tunable modulation frequency $f$ sets the macroscopic probing length through the thermal penetration depth $d_p$. The observed frequency dependence arises when $d_p$ sweeps across $d_{\text{NE}}$; the controlling parameter is therefore the ratio $d_p/d_{\text{NE}}$. This separation between microscopic excitation frequencies and experimentally accessible modulation frequencies is also consistent with prior thermoreflectance studies [22,27].

For water, $\omega_F$ is reported as $8.57 \times 10^{12}$ rad/s [28]. Using the vibrational density of states [29] and the standard heat-capacity integral [30,31], we partition the volumetric heat capacity into $C_1 = 0.18$ MJ m$^{-3}$ K$^{-1}$ and $C_2 = 4$ MJ m$^{-3}$ K$^{-1}$. This partitioning has only a minor influence on the fitted parameters (see Supplemental Material, Sec. 4); nevertheless, to be conservative we assign a 50% uncertainty to $C_1$. Simultaneous fitting across multiple frequencies yields $k_1 = (0.16 \pm 0.03)$ W m$^{-1}$ K$^{-1}$, $k_2 = (0.44 \pm 0.04)$ W m$^{-1}$ K$^{-1}$, $G_1 = 15^{48}_{-13}$ MW m$^{-2}$ K$^{-1}$, $G_2 = (45 \pm 12)$ MW m$^{-2}$ K$^{-1}$, and $g = (11.2 \pm 7.4) \times 10^9$ W m$^{-3}$ K$^{-1}$. The large uncertainty in $G_1$ reflects weak sensitivity in the solid-liquid ITC in this sample. In evaluating uncertainties, we account for propagation of known parameter errors, the effect of heat-capacity partitioning, and experimental noise. The apparent ITC ($G_A$) is defined as the ITC obtained by fitting dual-channel data with a single-channel model [12]. Using the fitted dual-channel parameters, $G_A$ reproduces the experimental frequency dependence in Fig. 2(a), confirming that the observed behavior originates from interfacial nonequilibrium transport.

As a cross-check, FDTR measurements were performed under identical conditions (see Supplemental Material, Fig. S5). Consistent with the SPS results, FDTR data cannot be fitted with a single constant ITC, but are quantitatively captured when the SPS-derived dual-channel parameters are used. Moreover, the decomposition of liquid thermal conductivity into potential–potential ($k^{\text{pp}}$), kinetic–kinetic ($k^{\text{kk}}$), and negligible cross ($k^{\text{pk}}$) contributions [32] further validates the model. For water, spectral data give $k^{\text{pp}} \approx 0.08$ W m$^{-1}$ K$^{-1}$ and $k^{\text{kk}} \approx 0.10$ W m$^{-1}$ K$^{-1}$ below $\omega_F$, yielding $k_1 = 0.18$ W m$^{-1}$ K$^{-1}$, in excellent agreement with our fitted value.

For octane, the situation differs slightly. Because $\omega_F$ has not been reliably reported, a rigorous frequency-resolved partition of heat capacity is not feasible. We therefore do not impose any a priori division of heat capacity between the two liquid channels. Instead, in the dual-channel fitting we fix the totals $C_1 + C_2$ and $k_1 + k_2$ to literature thermophysical values and their uncertainties, and fit the same five parameters as in the water case ($G_1$, $G_2$, $k_2$, $C_2$, $g$) under these hard constraints. Specifically, we fix $k_1 + k_2 = 0.1244$ W m$^{-1}$ K$^{-1}$, measured by the transient hot-wire method with a reported uncertainty of ~0.5% [33]. For the volumetric heat capacity at room temperature, literature values fall in a narrow range $C_1 + C_2 = 1.547$ - $1.57$ MJ m$^{-3}$ K$^{-1}$ [34,35]. We adopt $1.57$ MJ m$^{-3}$ K$^{-1}$ and assign a conservative 2% uncertainty that covers this spread. This strategy avoids speculative heat-capacity partitioning while keeping the model complexity identical.

The best-fit results are $G_1 = 3.5^{+\infty}_{-3.5}$ MW m$^{-2}$ K$^{-1}$, $G_2 = 8.2 \pm 5.9$ MW m$^{-2}$ K$^{-1}$, $k_2 = 0.1 \pm 0.07$ W m$^{-1}$ K$^{-1}$, $C_2 = 1.54 \pm 1.06$ MJ m$^{-3}$ K$^{-1}$, and $g = 3.2^{+4.8}_{-3.2} \times 10^9$ W m$^{-3}$ K$^{-1}$. The unbounded uncertainty in $G_1$ again reflects its negligible role in the fits. Notably, channel 2 accounts for ~98% of the total heat capacity, consistent with the exceptionally low $\omega_F$ inferred for octane and the long relaxation times reported in Ref. [36]. The apparent ITC derived from this model [Fig. 2(a)] agrees with experiment, supporting that the same minimal two-channel framework captures both liquids without introducing additional degrees of freedom.

As shown in Fig. 2(b), the modulation frequency is converted into the thermal penetration depth $d_p$ in the liquid, while the ITC is expressed in terms of the Kapitza length $l_k$. Remarkably, the Al-water and Al-octane data collapse onto a nearly identical curve, indicating comparable nonequilibrium lengths $d_{NE}$ and suggesting a universal scaling behavior independent of hydrogen bonding. The $l_k$-$d_p$ variation follows a logistic function, with an inflection point at $d_p \approx 3670$ nm. This value agrees well with the fitted nonequilibrium length for water ($d_{NE} \approx 3240$ nm), confirming the validity of the functional form. Similar to the frequency dependence of ITC, the $l_k$-$d_p$ relationship exhibits a characteristic S-shape that can be divided into three regimes. When $d_p < 0.1 d_{NE} \ll d_{NE}$, the two channels remain fully equilibrated, giving $G_A \approx G_1 + G_2$ and thus an almost constant $l_k$. As $d_p$ approaches $d_{NE}$, progressive decoupling occurs, leading to a rapid rise in $l_k$ that appears as a concave segment. For $d_p > d_{NE}$, the decoupling is essentially complete but continues weakly, resulting in a slower increase in $l_k$ that corresponds to a convex segment. These physical processes naturally give rise to the observed sigmoidal dependence of $l_k$ on $d_p$.

Through two representative case studies, we uncover that the coupling between the two heat-transfer modes in solid-liquid systems is extremely weak: the coupling coefficient is about three orders of magnitude smaller than that between optical and acoustic phonons in solids, and nearly seven orders of magnitude smaller than electron-phonon coupling in metals. In liquids, such weak coupling prevents the two modes from equilibrating on the same timescale, which directly gives rise to the frequency dependence of ITC. This discovery establishes new boundary conditions for modeling liquid interfacial heat transport.

Conventional extensions of the acoustic mismatch model (AMM) and diffuse mismatch model (DMM) for solid-liquid interfaces typically treat contributions from different modes as independently additive, which is effectively equivalent to assuming infinitely fast inter-mode equilibration [16,17]. Our results demonstrate instead that liquids must be treated within a multi-channel framework, where finite inter-mode coupling is explicitly included. This not only enables a more accurate description of the frequency response of ITC, but also provides experimental constraints for constructing a unified model of cross-phase energy transport.

This perspective also provides a bridge between simulation and pump-probe measurements. Nonequilibrium molecular dynamics (NEMD) simulations are commonly used to compute solid-liquid ITC from the steady-state interfacial temperature jump. That procedure corresponds to the quasi-static limit of thermoreflectance, i.e., the effective ITC as $f \to 0$, where the probing depth becomes large and internal nonequilibrium can strongly influence the apparent response. Although accessing this limit is experimentally challenging without broadband, high-SNR measurements, our data show that the apparent ITC decreases substantially toward low frequency. In other words, NEMD-predicted ITCs may be lower than transient thermoreflectance values measured at finite $f$, because the two approaches probe different points along the same frequency-dependent response curve. This interpretation is consistent with prior reports for the Si-water interface, where NEMD [37,38] yields a much larger Kapitza length than transient thermoreflectance measurements [39].

Finally, motivated by prior reports of direction-dependent interfacial heat transfer (thermal rectification) [40], we performed additional measurements with the heat-flow direction reversed across the water-Al interface (see Supplemental Material, Sec. 3.6, including Ref [39]). We find that the extracted vibration-dominated ITC is not identical in the two directions: $G_2$ changes from $(45 \pm 12)$ MW m$^{-2}$ K$^{-1}$ for Al-water to $(23 \pm 4)$ MW m$^{-2}$ K$^{-1}$ for water-Al, while the diffusional-channel ITC $G_1$ remains weakly constrained by sensitivity. Such rectification can arise when reversing the heat-flow direction changes the interfacial microscopic state and/or the efficiency of energy exchange with interfacial vibrational channels, as discussed in prior studies [41,42].

In summary, through precise measurements and rigorous data analysis, we report the first experimental observation of frequency-dependent solid-liquid ITC. We further demonstrate that this behavior originates from two distinct heat transport modes in the liquid, diffusional and phonon-like, that couple only weakly, in sharp contrast to the strong couplings typical of solid-state systems. Furthermore, we identify three distinct regimes of solid-liquid interfacial heat transport. Taken together, these discoveries establish a new framework for understanding nonequilibrium energy exchange at liquid interfaces.

The authors would like to express their sincere gratitude to Dr. Pan Li from the Center of Optoelectronic Micro & Nano Fabrication and Characterization Facility at the Wuhan National Laboratory for Optoelectronics, Huazhong University of Science and Technology, for his assistance with the electron-beam deposition process. This research was supported by the National Natural Science Foundation of China (NSFC) under Grant No. 52376058.